\documentclass[aoas,MSNbibl,nameyear,seceqn,dvips]{arximspdf}
\usepackage{graphicx}
\doi{10.1214/13-AOAS626} 
\volume{7}
\issue{2}
\pubyear{2013}
\firstpage{1139}
\lastpage{1161}

\makeatletter
\newcommand{\rrvert}{\vert}
\newcommand{\llvert}{\vert}
\makeatother

\begin{document}
\begin{frontmatter}

\title{Travel time estimation for ambulances using Bayesian data
augmentation\thanksref{T1}}
\thankstext{T1}{Supported in part by NSF Grant CMMI-0926814 and NSF
Grant DMS-12-09103.}
\runtitle{Travel time estimation using Bayesian data augmentation}

\begin{aug}
\author[A]{\fnms{Bradford~S.}~\snm{Westgate}\corref{}\ead[label=e1]{bsw62@cornell.edu}},
\author[A]{\fnms{Dawn~B.}~\snm{Woodard}\ead[label=u2,url]{http://people.orie.cornell.edu/woodard/}},
\author[B]{\fnms{David~S.}~\snm{Matteson}\ead[label=e3]{matteson@cornell.edu}}
\and
\author[A]{\fnms{Shane~G.}~\snm{Henderson}\ead[label=e4]{sgh9@cornell.edu}}
\runauthor{Westgate, Woodard, Matteson and Henderson}
\affiliation{Cornell University}
\address[A]{B.~S. Westgate\\
D.~B. Woodard\\
S.~G. Henderson\\
School of Operations Research\\
\quad and Information Engineering\\
Cornell University\\
Rhodes Hall\\
Ithaca, New York 14853\\
USA\\
\printead{e1}\\
\phantom{E-mail:\ }\printead*{e4}\\
\printead{u2}}

\address[B]{D.~S. Matteson\\
Department of Statistical Science\\
Cornell University\\
1196 Comstock Hall\\
Ithaca, New York 14853\\
USA\\
\printead{e3}}
\end{aug}

\received{\smonth{1} \syear{2012}}
\revised{\smonth{10} \syear{2012}}

%
\begin{abstract}
We introduce a Bayesian model for estimating the distribution of
ambulance travel times on each road segment in a city, using Global
Positioning System (GPS) data. Due to sparseness and error in the GPS
data, the exact ambulance paths and travel times on each road segment
are unknown. We simultaneously estimate the paths, travel times, and
parameters of each road segment travel time distribution using Bayesian
data augmentation. To draw ambulance path samples, we use a novel
reversible jump Metropolis--Hastings step. We also introduce two simpler
estimation methods based on GPS speed data.

We compare these methods to a recently published travel time estimation
method, using simulated data and data from Toronto EMS. In both cases,
out-of-sample point and interval estimates of ambulance trip times from
the Bayesian method outperform estimates from the alternative methods.
We also construct probability-of-coverage maps for ambulances. The
Bayesian method gives more realistic maps than the recently published
method. Finally, path estimates from the Bayesian method interpolate
well between sparsely recorded GPS readings and are robust to GPS
location errors.
\end{abstract}

%

\end{frontmatter}

\section{Introduction}
\label{secintro}
Emergency medical service (EMS) providers prefer to assign the closest
available ambulance to respond to a new emergency [\citet{dean08}].
Thus, it is vital to have accurate estimates of the travel time of each
ambulance to the emergency location. An ambulance is often assigned to
a new emergency while away from its base [\citet{dean08}], so the
problem is more difficult than estimating response times from several
fixed bases. Travel times also play a central role in positioning bases
and parking locations [\citet{brolapsem03,goldberg04,hend09}].
Accounting for variability in travel times can lead to considerable
improvements in EMS management [\citet{erkut08,ingolf08}]. We introduce
methods for estimating the distribution of travel times for arbitrary
routes on a municipal road network using historical trip durations and
vehicle Global Positioning System (GPS) readings. This enables
estimation of fastest paths in expectation between any two locations,
as well as estimation of the probability an ambulance will reach its
destination within a given time threshold.

Most EMS providers record ambulance GPS information; we use data from
Toronto EMS from 2007--2008. The GPS data include locations, timestamps,
speeds, and vehicle and emergency incident identifiers. Readings are
stored every 200 meters (m) or 240 seconds (s), whichever comes first.
The true sampling rate is higher, but this scheme minimizes data
transmission and storage. This is standard practice across EMS
providers, though the storage rates vary [\citet{mason05}]. In related
applications the GPS readings can be even sparser; \citet{lou09}
analyzed data from taxis in Tokyo in which GPS readings are separated
by 1--2 km or more.

\begin{figure}[b]

\includegraphics{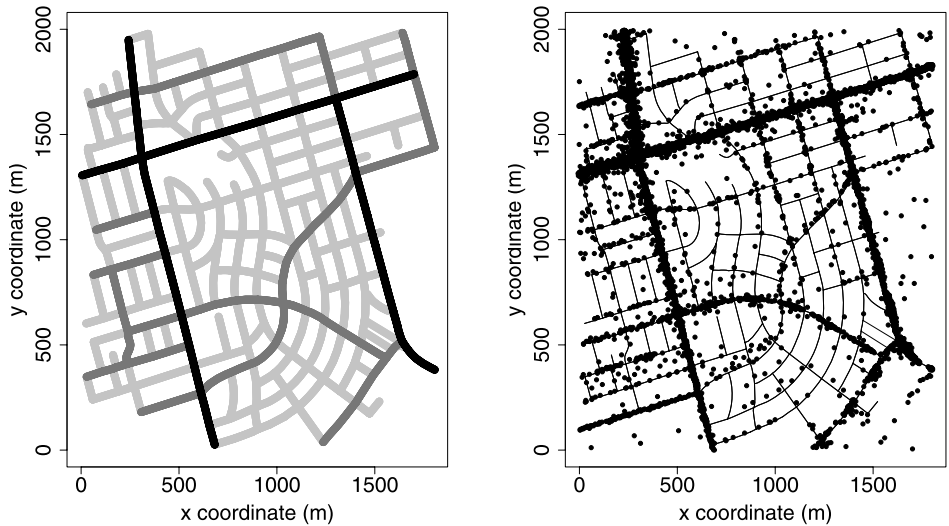}

\caption{Left: A subregion of Toronto, with primary roads (black),
secondary roads (gray)
and tertiary roads (light gray). Right: GPS data on this region from
the Toronto EMS lights-and-sirens data set.}
\label{figtestregion}
\end{figure}

The GPS location and speed data are also subject to error. Location
accuracy degrades in urban canyons, where GPS satellites may be
obscured and signals reflected [\citet{chen05,mason05,syed05}]. \citet
{chen05} observed average location errors of 27~m in parts of Hong Kong
with narrow streets and tall buildings, with some errors over 100~m.
Location error is also present in the Toronto data; see Figure \ref
{figtestregion}. \citet{witte04} found GPS speed errors of roughly 5\%
on average, with largest error at high speeds and when few GPS
satellites were visible.

Recent work on estimating ambulance travel time distributions has been
done by \citet{budge10} and \citet{aladdini10} using estimates based on
total trip distance and time, not GPS data. Budge et al. proposed
modeling the log travel times using a $t$-distribution, where the median
and coefficient of variation are functions of the trip distance (see
Section \ref{secbudge}). Aladdini found that the lognormal
distribution provided a good fit for ambulance travel times between
specific start and end locations. Budge et al. found heavier tails
than Aladdini, in part because they did not condition on the trip
location. Neither of these papers considered travel times on individual
road segments. For this reason they cannot capture some desired
features, such as faster response times to locations near major roads.

We first introduce two local methods using only the GPS locations and
speeds (Section \ref{secsimplemethods}). Each GPS reading is mapped
to the nearest road segment (the section of road between neighboring
intersections), and the mapped speeds are used to estimate the travel
time for each segment. In the first method, we use the harmonic mean of
the mapped GPS speeds to create a point estimator of the travel time.
We are the first to propose this estimator for mapped GPS data, though
it is commonly used for estimating travel times via speed data recorded
by loop detectors [\citet{rakha05,soriguera11,wardrop52}]. We give
theoretical results supporting this approach in the supplementary material [\citet
{west12supp}]. This method also yields interval and distribution
estimates of the travel time. In our second local method, we assume a
parametric distribution for the GPS speeds on each segment and
calculate maximum likelihood estimates of the parameters of this
distribution. These can be used to obtain point, interval, or
distribution estimates of the travel time.

In Sections \ref{secbayesian} and \ref{secgibbs}, we propose a more
sophisticated method, modeling the data at the trip level. Whereas the
local methods use only GPS data and the method of Budge et al. uses
only the trip start and end locations and times, this method combines
the two sources of information. We simultaneously estimate the path
driven for each ambulance trip and the distribution of travel times on
each road segment using Bayesian data augmentation [\citet{tanner87}].
For computation, we introduce a reversible jump Markov chain Monte
Carlo method [\citet{green95}]. Although parameter estimation is more
computationally intensive than for the other methods, prediction is
very fast. Also, the parameter estimates are updated offline, so the
increased computation time is not an operational handicap.

We compare the predictive accuracy on out-of-sample trips for the
Bayesian method, the local methods, and the method of Budge et al. on
a subregion of Toronto, using simulated data and real data (Sections
\ref{secsimexp} and \ref{secrealexp}). Since the methods have some
bias due in part to the GPS sampling scheme, we first use a
correction\vadjust{\goodbreak}
factor to make each method approximately unbiased (Section \ref
{secbias}). On simulated data, point estimates from the Bayesian
method outperform the alternative methods by over 50\% in root mean
squared error, relative to an Oracle method with the lowest possible
error. On real data, point estimates from the Bayesian method again
outperform the alternative methods. Interval estimates from the
Bayesian method have dramatically better coverage than intervals from
the local methods.

We also produce probability-of-coverage maps [\citet{budge10}], showing
the probability of traveling from a given intersection to any other
intersection within a time threshold (Section \ref{secthresh}). This
is the performance standard in many EMS contracts; an EMS organization
attempts to respond to, for example, 90\% of all emergencies within 9
minutes [\citet{fitch95}]. The estimates from the Bayesian method are
more realistic than those of Budge et al., because they differentiate
between equidistant locations based on whether or not they can be
reached by fast roads.

Finally, we assess the ambulance path estimates from the Bayesian
method (Sections \ref{secsimmapmatch} and \ref{secrealmapmatch}).
Estimating the path driven from a discrete set of GPS readings is
called the map-matching problem [\citet{mason05}]. Most map-matching
algorithms return a single path estimate [\citet{krumm07,lou09,marchal05,mason05}]. However, our posterior distribution can capture
multiple high-probability paths when the true path is unclear from the
GPS data. Our path estimates interpolate accurately between
widely-separated GPS locations and are robust to GPS error.

\section{Bayesian formulation}
\label{secbayesian}
\subsection{Model}
\label{secmodel}
Consider a network of $J$ directed road segments, called arcs, and a
set of $I$ ambulance trips on this network. Assume that each trip
starts and finishes on known nodes (intersections) $d_i^s$ and $d_i^f$
in the network, at known times $t_i^s$ and $t_i^f$. Therefore, the
total travel time $t_i^f - t_i^s$ is known. In practice, trips
sometimes begin or end in the interior of a road segment, however, road
segments are short enough that this is a minor issue. The median road
segment length in the full Toronto network is 111~m, the mean is 162~m,
and the maximum is 4613~m. Each trip $i$ has observed GPS readings,
indexed by $\ell\in\{ 1,\ldots, r_i \}$, and gathered at known times
$t_i^\ell$. GPS reading $\ell$ is the triplet $ (X_i^\ell, Y_i^\ell, V_i^\ell )$, where $X_i^\ell$ and $Y_i^\ell$ are the measured
geographic coordinates and $V_i^\ell$ is the measured speed. Denote
$G_i =  \{ (X_i^\ell, Y_i^\ell, V_i^\ell ) \}_{\ell=
1}^{r_i}$.

The relevant unobserved variables for each trip $i$ are the following:
\begin{longlist}[1.]
\item[1.] The unknown path (sequence of arcs) $A_i = \{A_{i,1},\ldots,A_{i,N_i}\}$ traveled by the ambulance from $d_i^s$ to $d_i^f$. The
path length $N_i$ is also unknown.
\item[2.] The unknown travel times $T_i = (T_{i,1}, \ldots,T_{i,N_i})$ on
the arcs in the path. We use the notation $T_i(j)$ to refer to the
travel time in trip $i$ on arc $j$.
\end{longlist}\eject

We model the observed and unobserved variables $ \{A_i, T_i,
G_i \}_{i=1}^I$ as follows. Conditional on $A_i$, each element
$T_{i,k}$ of the vector $T_i$ follows a lognormal distribution with
parameters $\mu_{A_{i,k}}, \sigma^2_{A_{i,k}}$, independently across
$i$ and $k$. We use the notation $T_{i,k} | A_i \sim\mathcal{LN}
(\mu_{A_{i,k}}, \sigma^2_{A_{i,k}} )$. In the literature,
ambulance travel times between specific locations have been observed
and modeled to be lognormal [\citet{aladdini10,alanis12}]. Denote the
expected travel time on each arc $j \in\{1,\ldots,J\}$ by $\theta(j) =
\exp (\mu_j + \sigma_j^2/2 )$. We use a multinomial logit
choice model [\citet{mcfadden73}] for the path $A_i$, with likelihood
%
\begin{equation}
f(A_i) = \frac{\exp (-C \sum_{k=1}^{N_i} \theta (A_{i,k}
) )}{\sum_{a_i \in\mathcal{P}_i} \exp (-C \sum_{k=1}^{n_i}
\theta (a_{i,k} ) )}, \label{eqnmulti}
\end{equation}
where $C>0$ is a fixed constant, $\mathcal{P}_i$ is the set of possible
paths with no repeated nodes from $d_i^s$ to $d_i^f$ in the network,
and $a_i = \{a_{i,1},\ldots,a_{i,n_i}\}$ indexes the paths in~$\mathcal
{P}_i$. In this model, the fastest routes in expectation have the
highest probability.

We assume that ambulances travel at constant speed on a single arc in a
given trip. This approximation is necessary since there is typically at
most one GPS reading on any arc in a given trip, and thus little
information in the data regarding changes in speed on individual arcs.
Therefore, the true location and speed of the ambulance at time
$t_i^\ell$ are deterministic functions $\operatorname{loc} (A_i,T_i,
t_i^\ell )$ and $\operatorname{sp} (A_i,T_i, t_i^\ell )$ of $A_i$
and $T_i$. Conditional on $A_i,T_i$, the measured location $
(X_i^\ell,Y_i^\ell )$ is assumed to have a bivariate normal
distribution [a standard assumption; see \citet{krumm07,mason05}]
centered at $\operatorname{loc} (A_i,T_i,t_i^\ell )$, with known
covariance matrix $\Sigma$. Similarly, the measured speed $V_i^\ell$ is
assumed to have a lognormal distribution with expectation equal to
$\operatorname{sp} (A_i,T_i, t_i^\ell )$ and variance parameter $\zeta^2$:
%
\begin{eqnarray}
\label{eqngps}  &\bigl(X_i^\ell,
Y_i^\ell \bigr) \big\vert A_i, T_i
\sim N_2 \bigl(\operatorname{loc} \bigl(A_i,T_i,t_i^\ell
\bigr), \Sigma \bigr),&
\\
&\displaystyle \log V_i^\ell\Big\vert A_i,
T_i \sim N \biggl( \log\operatorname{sp} \bigl(A_i,T_i,t_i^\ell
\bigr) - \frac{\zeta^2}{2}, \zeta^2 \biggr). &\label{eqngps2}
\end{eqnarray}
We assume independence between all the GPS speed and location errors.
Combining equations (\ref{eqnmulti})--(\ref{eqngps2}), we obtain the likelihood
%
\begin{eqnarray}
\label{eqnlikeli}
&& f \bigl( \{A_i,T_i,G_i
\}_{i=1}^I  \big\vert \bigl\{ \mu_j,
\sigma_j^2 \bigr\}_{j=1}^J,
\zeta^2 \bigr)
\nonumber
\\
&&\qquad= \prod_{i=1}^I
\Biggl[ f(A_i) \prod_{k=1}^{N_i}
\mathcal{LN} \bigl( T_{i,k}; \mu_{A_{i,k}}, \sigma^2_{A_{i,k}}
\bigr)
\nonumber
\\[-8pt]
\\[-8pt]
\nonumber
&&\hspace*{51pt}{}\times  \prod_{\ell=1}^{r_i} \biggl[
N_2 \bigl( \bigl(X_i^\ell,
Y_i^\ell \bigr); \operatorname{loc} \bigl(A_i,T_i,t_i^\ell
\bigr), \Sigma \bigr) \\
&&\hspace*{82pt}{}\times \mathcal{LN} \biggl( V_i^\ell;
\log\operatorname{sp} \bigl(A_i,T_i,
t_i^\ell \bigr) - \frac{\zeta^2}{2}, \zeta^2
\biggr) \biggr] \Biggr].\nonumber
\end{eqnarray}
In practice, we use data-based choices for the constants $\Sigma$ and
$C$ (see the supplementary material [\citet
{west12supp}]). The unknown parameters in the model are the
arc travel time parameters $ \{\mu_j, \sigma_j^2 \}_{j=1}^J$
and the GPS speed error parameter~$\zeta^2$.

\subsection{Prior distributions}
\label{secpriors}
To complete the model, we specify independent prior distributions for
the unknown parameters. We use $\mu_j \sim N (m_j, s^2 )$,
$\sigma_j \sim\operatorname{Unif} (b_1, b_2 )$, and $\zeta\sim\operatorname
{Unif} (b_3, b_4 )$,
where $m_j$, $ s^2$, $b_1$, $b_2$, $b_3$, $b_4$ are fixed
hyperparameters. A normal prior is a standard choice for the location
parameter of a lognormal distribution. We use uniform priors on the
standard deviations $\sigma_j$ and $\zeta$ [\citet{gelman06}]. The prior
ranges $[b_1,b_2]$ and $[b_3,b_4]$ are made wide enough to capture all
plausible parameter values. The prior mean for $\mu_j$ depends on $j$,
because there are often existing road speed estimates that can be used
to specify $m_j$. Prior information regarding the values $s^2$, $b_1$,
$b_2$, $b_3$, $b_4$ is more limited. We use a combination of prior
information and the data to specify all hyperparameters, as described
in the supplementary material [\citet
{west12supp}].

\section{Bayesian computational method}
\label{secgibbs}
We use a Markov chain Monte Carlo method to obtain samples\vspace*{1pt} $ (\zeta
^{2(\ell)},  \{\mu_j^{(\ell)}, \sigma_j^{(\ell)} \}_{j=1}^J,
\{A_i^{(\ell)}, T_i^{(\ell)} \}_{i=1}^{I}  )$ from the
joint posterior distribution of all unknowns [\citet{robert04,tierney94}]. Each unknown quantity is updated in turn, conditional on
the other unknowns, via either a draw from the closed-form conditional
posterior distribution or a Metropolis--Hastings (M--H) move. Estimation
of any desired function $g (\zeta^2,  \{\mu_j, \sigma_j^2
\}_{j=1}^J )$ of the unknown parameters is done via Monte Carlo,
taking $\hat{g} = \frac{1}{M} \sum_{\ell= 1}^M g (\zeta^{2(\ell)},
\{\mu_j^{(\ell)}, \sigma_j^{2(\ell)} \}_{j=1}^J ).$

\subsection{Markov chain initial conditions}
\label{secgibbsinitial}
To initialize each path $A_i$, select the middle GPS reading, reading
number $\lfloor r_i / 2 \rfloor+ 1$. Find the nearest node in the road
network to this GPS location, and route the initial path $A_i$ through
this node, taking the shortest-distance path to and from the middle
node. To initialize the travel time vector $T_i$, distribute the known
trip time across the arcs in the path $A_i$, weighted by arc length.
Finally, to initialize $\zeta^2$ and each $\mu_j$ and $\sigma^2_j$,
draw from their priors.

\subsection{Updating the paths}
\label{secAprop}
Updating the path $A_i$ may also require updating the travel times
$T_i$, since the number of arcs in the path may change. Since this
changes the dimension of the vector $T_i$, we update $(A_i,T_i)$ using
a reversible jump M--H move [\citet{green95}]. Calling the current values
$ (A_i^{(1)}, T_i^{(1)} )$, we propose new values $
(A_i^{(2)}, T_i^{(2)} )$ and accept them with the appropriate
probability, detailed below.

The proposal changes a contiguous subset of the path. The length
(number of arcs) of this subpath is limited to some maximum value $K$;
we specify $K$ in Section~\ref{secconv}. Precisely:
\begin{longlist}[3.]
\item[1.] With equal probability, choose a node $d^\prime$ from the path
$A_i^{(1)}$, excluding the final node.
\item[2.] Let $a^{(1)}$ be the number of nodes that follow $d^\prime$ in
the path. With equal probability, choose an integer $w \in\{1, \ldots,
\min (a^{(1)}, K )\}$. Denote the $w$th node following
$d^\prime$ as $d^{\prime\prime}$. The subpath from $d^\prime$ to
$d^{\prime\prime}$ is the section to be updated (the ``current update
section'').
\item[3.] Consider all possible routes of length up to $K$ from $d^\prime$
to $d^{\prime\prime}$. With equal probability, propose one of these
routes as a change to the path (the ``proposed update section''),
giving the proposed path $A_i^{(2)}$.
\end{longlist}

Next we propose travel times $T^{(2)}_i$ that are compatible with
$A_i^{(2)}$. Let $\{c_1, \ldots,  c_m\} \subset A_i^{(1)}$ and $\{p_1,
\ldots, p_n\} \subset A_i^{(2)}$ denote the arcs in the current and
proposed update sections, noting that $m$ and $n$ may be different.
Recall that $T_i(j)$ denotes the travel time of trip $i$ on arc $j$.
For each arc $j \in A_i^{(2)} \setminus\{p_1, \ldots, p_n\}$, set
$T^{(2)}_i(j) = T_i^{(1)}(j)$. Let $S_i = \sum_{\ell=1}^m
T_i^{(1)}(c_\ell)$ be the total travel time of the current update
section. Since the total travel time of the entire trip is known (see
Section~\ref{secmodel}), $S_i$ is fixed and known as well, conditional
on the travel times for the arcs that are unchanged by this update.
Therefore, we must have $\sum_{\ell=1}^n T^{(2)}_i(p_\ell) = S_i$. The
travel times $T^{(2)}_i (p_1), \ldots, T^{(2)}_i (p_n)$ are proposed by
drawing $(r_1, \ldots, r_n) \sim\operatorname{Dirichlet} (\alpha\theta
(p_1), \ldots, \alpha\theta(p_n) )$ for a constant $\alpha> 0$
(specified below), and setting $T^{(2)}_i(p_\ell) = r_\ell S_i$ for
$\ell\in\{1,\ldots,n\}$. The expected value of the proposed travel
time on arc $p_\ell$ is $E (T^{(2)}_i(p_\ell) ) = S_i \frac
{\theta(p_\ell)}{\sum_{k=1}^n \theta(p_k)}$. Therefore,\vspace*{1pt} the expected
values of the proposed times are weighted by the arc travel time
expected values [\citet{gelman04}]. The constant $\alpha$ controls the
variances and covariances of the components $T_i^{(2)}(p_\ell)$. In our
experience $\alpha= 1$ works well; one can also tune $\alpha$ to
obtain a desired acceptance rate for a particular data set [\citet
{robert04,roberts01}].

Let $N_i^{(j)}$ be the number of edges in the path $A_i^{(j)}$ for $j
\in\{1,2\}$, and let $a^{(2)}$ be the number of nodes that follow
$d^\prime$ in the path $A_i^{(2)}$. We accept the proposal $(A^{(2)}_i,
T_i^{(2)})$ with probability equal to the minimum of one and
%
\begin{eqnarray}\label{EqnAccProb}
&& \frac{f_i (A^{(2)}_i, T^{(2)}_i, G_i \vert  \{\mu_j,
\sigma^2_j
\}_{j=1}^{J}, \zeta^2 )}{f_i (A_i^{(1)},
T_i^{(1)}, G_i
\vert  \{\mu_j, \sigma^2_j  \}_{j=1}^{J}, \zeta^2
)}\times\frac{N_i^{(1)} \min(a^{(1)},K)}{N^{(2)}_i \min
(a^{(2)}, K)}
\nonumber
\\[-8pt]
\\[-8pt]
\nonumber
& &\qquad \times\frac{\operatorname{Dir} ({T_i^{(1)}(c_1)}/{S_i}, \ldots,
{T_i^{(1)}(c_m)}/{S_i}; \alpha\theta(c_1),\ldots,
\alpha\theta(c_m) )}{\operatorname{Dir} ({T_i^{(2)}(p_1)}/{S_i},
\ldots,
{T_i^{(2)}(p_n)}/{S_i};
\alpha\theta(p_1), \ldots, \alpha\theta(p_n) )}S_i^{n-m},
\end{eqnarray}
where $f_i$ is the contribution of trip $i$ to equation (\ref
{eqnlikeli}) and $\operatorname{Dir}(x; y)$ denotes the Dirichlet density with
parameter vector $y$, evaluated at $x$. The proposal density for the
travel times $T^{(2)}_i (p_1), \ldots, T^{(2)}_i (p_n)$ requires a
change of variables from the Dirichlet density. This leads to the
factor $S_i^{n-m}$ in the ratio of proposal densities. In the supplementary material [\citet
{west12supp}], we show that this move is valid since it is reversible with
respect to the conditional posterior distribution of $(A_i, T_i)$.

\subsection{Updating the trip travel times}
To update the realized travel time vector $T_i(j)$, we use the
following M--H move. Given current travel times $T_i^{(1)}$, we propose
travel times $T_i^{(2)}$:
\begin{longlist}[2.]
\item[1.] With equal probability, choose a pair of distinct arcs $j_1$ and
$j_2$ in the path $A_i$. Let $S_i = T^{(1)}_i(j_1)+T^{(1)}_i(j_2)$.
\item[2.] Draw $(r_1, r_2) \sim\operatorname{Dirichlet} (\alpha^\prime\theta
(j_1), \alpha^\prime\theta(j_2) )$. Set $T^{(2)}_i(j_1) = r_1S_i$
and\break $T^{(2)}_i(j_2) = r_2S_i$.
\end{longlist}
Similarly to the path proposal above, this proposal randomly
distributes the travel time over the two arcs, weighted by the expected
travel times $\theta(j_1)$ and $\theta(j_2)$, with variances controlled
by the constant $\alpha^\prime$ [\citet{gelman04}]. In our experience
$\alpha^\prime= 0.5$ is effective for our application. It is
straightforward to calculate the M--H acceptance probability.

\subsection{\texorpdfstring{Updating the parameters $\mu_j$, $\sigma_j^2$, and $\zeta^2$}
{Updating the parameters mu j, sigma j2, and zeta 2}}
To update each $\mu_j$, we sample from the full conditional posterior
distribution, which is available in closed form. We have
$\mu_j \llvert \sigma^2_j,  \{A_i,T_i \}_{i=1}^I \sim N
(\hat{\mu}_j, \hat{s}^2_j )$,
where
\[
\hat{s}^2_j = \biggl[ \frac{1}{s^2} +
\frac{n_j}{\sigma^2_j} \biggr]^{-1},\qquad \hat{\mu}_j =
\hat{s}^2_j \biggl[ \frac{m_j}{s^2} +
\frac{1}{\sigma^2_j} \sum_{i \in I_j} \log T_i(j)
\biggr],
\]
the set $I_j \subset\{1, \ldots, I\}$ indicates the subset of trips
using arc $j$, and $n_j = \llvert  I_j \rrvert $.

To update each $\sigma^2_j$, we use a local M--H step [\citet
{tierney94}]. We propose $\sigma_j^{2*} \sim\mathcal{LN}( \log\sigma
_j^2, \eta^2)$, having fixed variance $\eta^2$. The M--H acceptance
probability $p_{\sigma}$ is the minimum of 1 and
\[
\frac{\sigma_j}{\sigma^{*}_j}\mathbf{1}_{\{\sigma^{*}_j \in[b_1, b_2]
\}} \biggl(\frac{\prod_{i \in I_j} \mathcal{LN} (T_i(j); \mu_j, \sigma
^{2*}_j )}{\prod_{i \in I_j} \mathcal{LN} (T_i(j); \mu_j,
\sigma^2_j )} \biggr)
\frac{\mathcal{LN} (\sigma^2_j; \log (\sigma^{2*}_j ), \eta
^2 )}{\mathcal{LN} (\sigma^{2*}_j; \log (\sigma^2_j
), \eta^2 )}.
\]

To update $\zeta^2$, we use another M--H step with a lognormal proposal,
with variance $\nu^2$. The proposal variances $\eta^2, \nu^2$ are tuned
to achieve an acceptance rate of approximately 23\% [\citet{roberts01}].

\subsection{Markov chain convergence}
\label{secconv}
The transition kernel for updating the path $A_i$ is irreducible, and
hence valid [\citet{tierney94}], if it is possible to move between any
two paths in $\mathcal{P}_i$ in a finite number of iterations, for all
$i$. For a given road network, the maximum update section length $K$
can be set high enough to meet this criterion. However, the value of
$K$ should be set as low as possible, because increasing $K$ tends to
lower the acceptance rate. If there is a region of the city with sparse
connectivity, the required value of $K$ may be impractically large. For
example, there could be a single arc of a highway alongside many arcs
of a parallel minor road. Then, a large $K$ would be needed to allow
transitions between the highway and the minor road. If $K$ is kept
smaller, the Markov chain is reducible. In this case, the chain
converges to the posterior distribution restricted to the closed
communicating class in which the chain is absorbed. If this class
contains much of the posterior mass, as might arise if the initial path
follows the GPS data reasonably closely, then this should be a good
approximation.

In Sections \ref{secsimexp} and \ref{secrealexp}, we apply the
Bayesian method to simulated data and data from Toronto EMS, on a
subregion of Toronto with 623 arcs. Each Markov chain was run for
50,000 iterations (where each iteration updates all parameters), after
a burn-in period of 25,000 iterations. We calculated Gelman--Rubin
diagnostics [\citet{gelman92}], using two chains, for the parameters
$\zeta^2$, $\mu_j$, and $\sigma^2_j$. Results from a typical simulation
study were as follows: potential scale reduction factor of 1.06 for
$\zeta^2$, of less than 1.1 for $\mu_j$ for 549 arcs (88.1\%), between
1.1--1.2 for 43 arcs (6.9\%), between 1.2--1.5 for 30 arcs (4.8\%), and
less than 2 for the remaining 1 arc, with similar results for the
parameters $\sigma^2_j$. These results indicate no lack of convergence.

Each Markov chain run for these experiments takes roughly 2 hours on a
3.2~GHz workstation. Each iteration of the Markov chain scales linearly
in time with the number of arcs and the number of ambulance trips: $O(J
+ I)$, assuming the lengths of the ambulance paths do not grow as well.
This assumption is reasonable, since long ambulance paths are
undesirable for an EMS provider. It is much more difficult to assess
how the number of iterations required for convergence changes with $J$
and $I$, since this would require bounding the spectral gap of the
Markov chain. The full Toronto road network has roughly 110 times as
many arcs as the test region, and the full Toronto EMS data set has
roughly 80 times as many ambulance trips.

In practice, parameter estimates are updated infrequently and offline.
Once parameter estimation is done, prediction for new routes and
generation of our figures is very fast. If parameter estimation for the
Bayesian method is computationally impractical for the entire city, it
can be divided into multiple regions and estimated in parallel. We
envision creating overlapping regions and discarding estimates on the
boundary to eliminate edge effects (see Section \ref{secrealdata}).
During parameter estimation, trips traveling through multiple regions
would be divided into portions for each region, as we have done in our
Toronto EMS experiments. However, prediction for such a trip can be
handled directly, given the parameter estimates for all arcs in the
city. The fastest path in expectation may be calculated using a
shortest path algorithm over the entire road network, which gives a
point estimate of the trip travel time. A distribution estimate of the
travel time can be obtained by sampling travel times on the arcs in
this fastest path (see Section \ref{secrealcomp}).

\section{Comparison methods}
\subsection{Local methods}
\label{secsimplemethods}
Here we detail the two local methods outlined in Section~\ref
{secintro}. Each GPS reading is mapped to the nearest arc (both
directions of travel are treated together). Let $n_j$ be the number of
GPS points mapped to arc $j$, $L_j$ the length of arc $j$, and $ \{
V_j^k \}_{k=1}^{n_j}$ the mapped speed observations. We assume
constant speed on each arc, as in the Bayesian method. Thus, let $T_j^k
= L_j / V_j^k$ be the travel time associated with observed speed $V_j^k$.

In the first local method, we calculate the harmonic mean of the speeds
$ \{V_j^k \}_{k=1}^{n_j}$ and convert to a travel time point estimate
\[
\hat{T}_j^H = \frac{L_j}{n_j}\sum
_{k=1}^{n_j}\frac{1}{V^k_j}.
\]
This is equivalent to calculating the arithmetic mean of the associated
travel times~$T_j^k$. The empirical distribution of the associated
times $ \{T_j^k \}_{k=1}^{n_j}$ can be used as a distribution
estimate. Because readings with speed 0 occur in the Toronto EMS data
set, we set any reading with speed below 5 miles per hour (mph) equal
to 5~mph. This harmonic mean estimator is well known in the
transportation research literature, where it is called the ``space mean
speed,'' in the context of estimating travel times using speed data
recorded by loop detectors [\citet{rakha05,soriguera11,wardrop52}].

In the supplementary material [\citet
{west12supp}], we consider this travel time estimator $\hat{T}_j^H$
and its relation to the GPS sampling scheme.\vspace*{-1pt} We show that if GPS points
are sampled by distance (e.g., every 100~m), $\hat{T}_j^H$ is an
unbiased estimator for the true mean travel time.
However, if GPS points are sampled by time (e.g., every 30 s),
$\hat{T}_j^H$ overestimates the mean travel time. The Toronto EMS data
set uses a combination of sampling-by-distance and sampling-by-time.
However, the distance constraint is usually satisfied first (see Figure
\ref{figrealpost}, where the sampled GPS points are regularly
spaced). Thus, the travel time estimator $\hat{T}_j^H$ is appropriate.

In the second local method, we assume $V_j^k \sim\mathcal{LN}(m_j,
s_j^2)$, independently across $k$, for unknown travel time parameters
$m_j$ and $s_j^2$.\vspace*{1pt} This distribution on the travel speed implies that
the travel times also have a lognormal distribution: $T_j^k \sim
\mathcal{LN} (\log(L_j) - m_j, s^2_j )$.\vadjust{\goodbreak} We use the maximum
likelihood estimators (MLEs)
\[
\hat{m}_j = \frac{1}{n_j}\sum_{k=1}^{n_j}
\log \bigl(V^k_j \bigr),\qquad \hat{s}^2_j
= \frac{1}{n_j}\sum_{k=1}^{n_j} \bigl(
\log \bigl(V^k_j \bigr) - \hat{m}_j
\bigr)^2
\]
to estimate $m_j$ and $s^2_j$. Our point travel time estimator is
\[
\hat{T}^{\operatorname{MLE}}_j = E \bigl(T_j\rrvert
\hat{m}_j, \hat{s}^2_j \bigr) = \exp \biggl(
\log(L_j) - \hat{m}_j + \frac{\hat{s}_j^2}{2} \biggr).
\]
This second local method also provides a natural distribution estimate
for the travel times via\vspace*{1pt} the estimated lognormal distribution for
$T_j^k$. Correcting for zero-speed readings is again done by
thresholding, to avoid $\log(0)$.

Some small residential arcs have no assigned GPS points in the Toronto
EMS data set (see Figure \ref{figtestregion}). In this case, we use a
breadth-first search [\citet{nilsson98}] to find the closest arc in the
same road class that has assigned GPS points. The road classes are
described in Section \ref{secsimexp}; by restricting our search to
arcs of the same class, we ensure that the speeds are comparable.

\subsection{\texorpdfstring{Method of Budge et al}{Method of Budge et al.}}
\label{secbudge}
Budge, Ingolfsson and Zerom
(\citeyear{budge10}) introduced a travel time distribution estimation method
relying on trip distance. Since the exact path traveled is usually
unknown, the length of the shortest-distance path between the start and
end locations is used as a surrogate for the true travel distance. The
method relies on the model $t_i = m(d_i) \exp [c(d_i) \varepsilon
_i ]$, where $t_i$ and $d_i$ are the total time and distance for
trip $i$, $\varepsilon_i$ follows a $t$-distribution with $\tau$ degrees of
freedom, and $m(\cdot)$ and $c(\cdot)$ are unknown functions. In their
preferred method, they assume parametric expressions for the functions
$m(\cdot)$ and $c(\cdot)$, and estimate the parameters using maximum likelihood.

We implemented this parametric method and compared it to a related
binning method. In the binning method, we divide the ambulance trips
into bins by trip distance and fit a separate $t$-distribution to the log
travel times for each bin. We then linearly interpolate between the
quantiles of the travel time distributions for adjacent bins to
generate a travel time distribution estimate for a trip of any
distance. On simulated data, the parametric and binning methods perform
very similarly, while on real data the binning method slightly
outperforms the parametric method. Thus, we report only results of the
binning method in Sections \ref{secsimexp}--\ref{secrealexp}.

\section{Bias correction}
\label{secbias}
We use a bias correction factor to make each method approximately
unbiased, because we have found that this improves performance for all
methods. There are several reasons why the methods result in biased
estimates, some inherent to the methods themselves and some due to
sampling characteristics of the GPS data. One source of bias is the
inspection paradox in the GPS data, discussed at length in the supplementary material [\citet
{west12supp}]. The Bayesian method is also biased because of the difference in
path estimation from the training to the test data. On the training
data, the Bayesian method uses the GPS data to estimate a solution to
the map-matching problem. On the test data, the estimated fastest path
between the start and end nodes is used to imitate the prediction
scenario where the route is not known beforehand. This leads to
underestimation of the true travel times.

Most commonly, bias correction is done using an asymptotic expression
for the bias [\citet{breslow95,kan09}]. We use an empirical bias
correction factor, because there is no analytic expression available.
The bias correction factor for each method is calculated in the
following manner. We divide the set of trips from each data set
randomly into training, validation, and test sets [\citet{hastie05}]. We
fit the methods on the training data, calculate a bias correction
factor on the validation data, and predict the travel times for the
trips in the test data. The data are split into 50\% training and 50\%
validation and test. To use the validation/test data most efficiently,
we do cross-validation: divide the validation/test data into ten sets,
use nine sets for the validation data, the tenth for the test data, and
repeat for all ten cases. For a given validation set of $n$ trips,
where the estimated trip travel times are $\{\hat{t}_i\}_{i=1}^n$ and
the true travel times are $\{t_i\}_{i=1}^n$, the bias correction factor is
\[
b = \frac{1}{n} \Biggl(\sum_{i=1}^n
\log\hat{t}_i - \sum_{i=1}^n
\log t_i \Biggr).
\]
Subtracting this factor from the log estimates on the test data makes
each method unbiased on the log scale. We calculate the bias correction
on the log scale because it is more robust to travel time outliers.

\section{Simulation experiments}
\label{secsimexp}
Next we test the Bayesian method, local methods, and the method of
Budge et al. on simulated data. We compare the accuracy of the four
methods for predicting travel times of test trips. We simulate
ambulance trips on the road network of Leaside, Toronto, shown in
Figure \ref{figtestregion} (roughly 4 square kilometers). This region
has four road classes; we define the highest-speed class to be primary
arcs, the two intermediate classes to be secondary arcs, and the
lowest-speed class to be tertiary arcs (Figure \ref{figtestregion}).
In the Leaside region, a value $K=6$ (see Section \ref{secconv})
guarantees that the Markov chain is valid.

\subsection{Generating simulated data}
\label{secgensimdata}
We simulate ambulance trips with true paths, travel times, and GPS
readings. For each trip $i$, we uniformly choose start and end nodes.
We construct the true path $A_i$ arc-by-arc. Beginning at the start
node, we uniformly choose an adjacent arc from those that lower the
expected time to the end node, and repeat until the end node is
reached. This method differs from the Bayesian prior (see Section \ref
{secmodel}) and can lead to a wide variety of paths traveled between
two nodes.

The arc travel times are lognormal: $T_{i,k} \sim\mathcal{LN}(\mu
_{A_{i,k}}, \sigma^2_{A_{i,k}})$. To set the true travel time
parameters $ \{\mu_j, \sigma^2_j \}$ for arc $j$, we uniformly
generate a speed between 20--40 mph. We draw $\sigma_j \sim\operatorname
{Unif} (0.5\log (\sqrt{3} ), 0.5\log(3) )$ and set
$\mu_j$ so that the arc length divided by the mean travel time equals
the random speed. The range for $\sigma_j$ generates a wide variety of
arc travel time variances. Comparisons between the estimation methods
are invariant to moderate changes in the $\sigma_j$ range.

We simulate data sets with two types of GPS data: good and bad. The
good GPS data sets are designed to mimic the conditions of the Toronto
EMS data set. Each GPS point is sampled at a travel distance of 250~m
after the previous point. Straight-line distance between GPS readings
is typically 200~m in the Toronto EMS data, but we simulate data via
the longer along-path distance. The GPS locations are drawn from a
bivariate normal distribution with $\Sigma= \bigl(
{ {100 \atop 0} \enskip {0 \atop 100}}
\bigr)$. The GPS speeds are drawn from a lognormal distribution with
$\zeta^2 = 0.004$, which gives a mean absolute error of 5\% of speed,
approximately the average result seen by \citet{witte04}.\looseness=-1

The bad GPS data sets are designed to be sparse and have GPS error
consistent with the high error results seen by \citet{chen05} and \citet
{witte04}. GPS points are sampled every 1000~m. The constant $\Sigma=
\bigl(
{ {465 \atop 0} \enskip{ 0 \atop 465}
}
\bigr)$, which gives mean distance of 27~m between the true and
observed locations, the average error seen in Hong Kong by \citet
{chen05}. The parameter $\zeta^2 = 0.01575$, corresponding to mean
absolute error of 10\% of speed, which is approximately the result from
low-quality GPS settings tested by \citet{witte04}.

\subsection{Travel time prediction}
\label{secsimpredict}
We simulate ten good GPS data sets and ten bad GPS data sets, as
defined above, each with a training set of 2000 trips and a
validation/test set of 2000 trips. Taking the true path for each test
trip as known and using the cross-validation approach of Section \ref
{secbias} to estimate bias correction factors, we calculate point and
95\% predictive interval estimates for the test set travel times using
the four methods. To obtain a gold standard for performance, we
implement an Oracle method. In this method, the true travel time
parameters $ \{\mu_j, \sigma^2_j \}$\vspace*{1pt} for each arc $j$ are
known. The true expected travel time for each test trip is used as a
point estimate. This implies that the Oracle method has the lowest
possible root mean squared error (RMSE) for realized travel time estimation.

We compare the predictive accuracy of the point estimates from the four
methods via the RMSE (in seconds), the RMSE of the log predictions
relative to the true log times (``RMSE log''), and the mean absolute
bias on the log scale over the test sets of the cross-validation
procedure (``Bias M.A.''). We calculate metrics on the log scale
because the residuals on the log scale are much closer to normally
distributed. On the original scale, there are several outlying trips in
the Toronto EMS data (Section \ref{secrealexp}) with very large
travel times that heavily influence the metrics. The bias metric
measures how well the bias correction works. If $k \in\{1,\ldots,10\}$
indexes the cross-validation test sets, where test set $k$ has $n_k$
trips with true travel times $t_{i}^{(k)}$ and estimates $\hat
{t}_{i}^{(k)}$, for $i \in\{1,\ldots,n_k\}$, then
%
\begin{equation}
\label{eqnbias} \operatorname{Bias (M.A.)} = \frac{1}{10}\sum
_{k=1}^{10} \Biggl\llvert \frac{1}{n_k} \Biggl(
\sum_{i=1}^{n_k} \log\hat{t}_i^{(k)}
- \sum_{i=1}^{n_k} \log t_i^{(k)}
\Biggr) \Biggr\rrvert.
\end{equation}

We compare the interval estimates using the the percentage of 95\%
predictive intervals that contain the true travel time (``Cov. \%'')
and the geometric mean width of the 95\% predictive intervals
(``Width''). Table \ref{tabsimoutofsample} gives arithmetic means
for these metrics over the ten good and bad simulated data
sets.\looseness=-1
%
\begin{table}
\caption{Out-of-sample trip travel time estimation performance on
simulated data}\label{tabsimoutofsample}
\begin{tabular*}{\textwidth}{@{\extracolsep{\fill}}lccccc@{}}
\hline
\multicolumn{1}{@{}l}{\textbf{Estimation method}} & \textbf{RMSE (s)} & \textbf{RMSE log} & \textbf{Bias (M.A.)} & \textbf{Cov. \%} &
\textbf{Width (s)}\\
\hline
\multicolumn{6}{c}{Good GPS data (Mean over ten data sets)} \\
Oracle & 15.9 & 0.183 & 0.010 & -- & -- \\
Bayesian & 16.1 & 0.187 & 0.010 & 95.8 & 57.2 \\
Local MLE & 16.8 & 0.196 & 0.010 & 94.4 & 56.8 \\
Local harm. & 16.8 & 0.196 & 0.010 & 94.0 & 56.2 \\
Budge et al. & 17.3 & 0.201 & 0.011 & 96.2 & 67.2 \\[6pt]
\multicolumn{6}{c}{Bad GPS data (Mean over ten data sets)} \\
Oracle & 16.4 & 0.183 & 0.012 & -- & -- \\
Bayesian & 16.9 & 0.191 & 0.013 & 96.1 & 60.4 \\
Local MLE & 18.1 & 0.209 & 0.014 & 92.3 & 57.8 \\
Local harm. & 18.1 & 0.209 & 0.014 & 90.9 & 55.5 \\
Budge et al. & 17.9 & 0.201 & 0.013 & 96.2 & 68.2 \\
\hline
\end{tabular*}
\end{table}

In both data set types, the point estimates from the Bayesian method
greatly outperform the estimates from the local methods and the method
of Budge et al. The Bayesian estimates closely approach the Oracle
estimates, especially on the good GPS data sets. In the good data sets,
the Bayesian method has 70\% lower error than the local methods in RMSE
on the log scale, and 78\% lower error than Budge et al., after
eliminating the unavoidable error of the Oracle method. In the bad data
sets, the Bayesian method outperforms the local methods by 70\% and
Budge et al. by 56\% in log scale RMSE, relative to the Oracle method.
The method of Budge et al. outperforms the local methods on the bad
GPS data, while the reverse holds for the good GPS data. 

\begin{figure}

\includegraphics{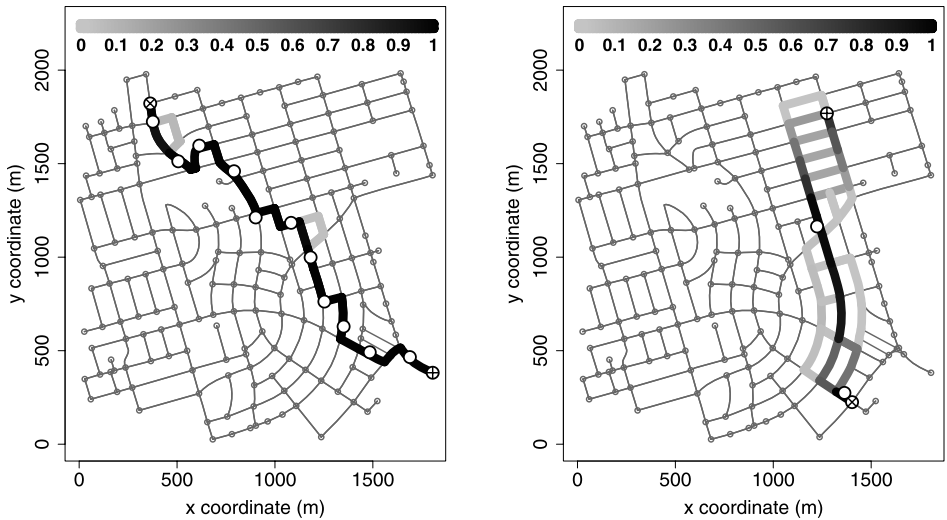}

\caption{Map-matching estimates for two simulated trips, shaded by the
probability each arc is traversed.}
\label{figsimpost}
\end{figure}

The Bayesian method also outperforms the other methods in interval
estimates. For the good GPS data, the interval estimates from the
Bayesian and local methods are similar, while the estimates from the
method of Budge et al. are substantially wider, with slightly higher
coverage percentage. For the bad GPS data, the intervals from the
Bayesian method have higher coverage percentage than the intervals from
the local methods, and the intervals from the method of Budge et al. are again wider, with no corresponding increase in coverage percentage.

\subsection{Map-matched path results}
\label{secsimmapmatch}
Next we assess path estimates from the Bayesian method for
representative paths, shown in Figure \ref{figsimpost}. The GPS
locations are shown in white. The starting node is marked with a cross
and the ending node with an X. Each arc is shaded in gray by the
marginal posterior probability that it is traversed in the path. Arcs
with probability less than 1\% are unshaded. The left-hand path is from
a good GPS data set, as defined in Section \ref{secgensimdata}. The
Bayesian method easily identifies the correct path. Every correct arc
has close to 100\% probability, and only two incorrect detours have
probability above 1\%. This is typical performance for trips with
good\vadjust{\goodbreak}
GPS data. The right-hand path is from a bad GPS data set. The sparsity
in GPS readings makes the path very uncertain. Near the beginning of
the path, there are five routes with similar expected travel times, and
the GPS readings do not distinguish between them, so each has roughly
20\% posterior probability. The Bayesian method is very effective at
identifying alternative routes when the true path is unclear.

\section{Analysis of Toronto EMS data}
\label{secrealexp}
Next we compare the four methods on the Toronto EMS data.
\subsection{Data}
\label{secrealdata}
The Toronto data consist of GPS data and trip information for ambulance
trips with one of two priority levels: lights-and-sirens (L--S) or
standard travel (Std). We address these separately, again focusing on
the Leaside subregion of Toronto. The right plot in Figure \ref
{figtestregion} shows the GPS locations for the L--S data set. This
data set contains 1930 ambulance trips and roughly 14,000 GPS points.
The primary roads tend to have a large amount of data, the secondary
roads a moderate amount, and the tertiary roads a small amount. The Std
data set is larger (3989 trips), with a similar spatial distribution of points.

We use only the portion of trips where the ambulance was driving to the
scene of an emergency, and discard trips for which this portion cannot
be identified. We also discard some trips (roughly 1\%) that would
impair estimation, for example, trips where the ambulance turned around
or where the ambulance stopped for a long period, not at a stoplight or
in traffic. Finally, most of the trips in the data set do not begin or
end in the subregion, they simply pass through, so we use the closest
node to the first GPS location as the approximated start node, and the
time of the first GPS reading as the start time. Similarly, we use the
last GPS reading for the end node. This produces some inaccuracy of
estimated travel times on the boundary of the region. This could be
fixed by applying our method to overlapping regions and discarding
estimates on the boundary.

\subsection{Arc travel time estimates}
\label{secrealspeeds}
\begin{figure}

\includegraphics{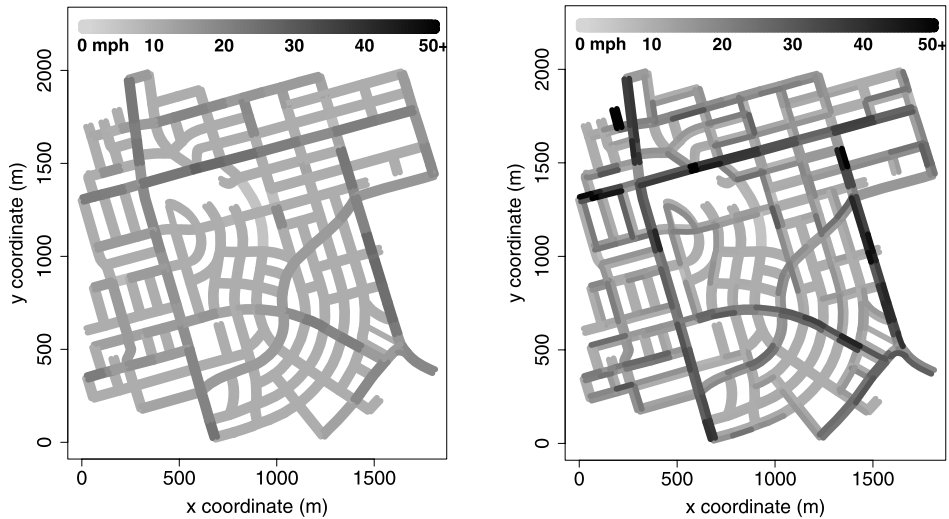}

\caption{Prior (left) and posterior (right) speeds from the Bayesian
method, for Toronto L--S data, in miles per hour (mph).}
\label{figreallsspeeds}
\end{figure}

Here we report the travel time estimates from the Bayesian method.
Toronto EMS has existing estimates of the travel times, which we use to
set the prior $ \{m_j \}_{j=1}^J$ hyperparameters\vspace*{1pt} (see the supplementary material [\citet
{west12supp}]). These estimates are different for L--S and Std trips, but are
the same for the two travel directions of parallel arcs. We have also
tested the Bayesian method with the data-based hyperparameters
described in the supplementary material [\citet
{west12supp}] and have observed similar performance.
Figure \ref{figreallsspeeds} shows prior and posterior speed
estimates (length divided by mean travel time) from the Bayesian method
on the L--S data set. Each arc is shaded in gray based on its speed
estimate, so most roads have two shades in the right-hand plot,
corresponding to travel in each direction.

The posterior speed estimates from the Bayesian method are reasonable;
primary arcs tend to have high speed estimates, and estimated speeds
for consecutive arcs on the same road are typically similar. Arcs
heading into major intersections (intersections between two primary or
secondary roads, as shown in Figure \ref{figtestregion}) are often
slower than the reverse arcs. In the corresponding figure for Std data
(not shown), the slowdown into major intersections is even more
pronounced. For most arcs the posterior estimate of the speed is higher
than the prior estimate, suggesting that the existing road speed
estimates used to specify the prior are underestimates.
%

There are a few arcs that have poor estimates from the Bayesian method.
For example, parallel black arcs in the top-left corner have poor
estimates due to edge effects. Also, some short interior arcs have
unrealistically high estimates, likely because there are few GPS points
on these arcs. This undesirable behavior could be reduced or eliminated
by using a random effect prior distribution [\citet{gelman04}] for roads
in the same class, which would have the effect of pooling the available data.

\subsection{Travel time prediction}
\label{secrealcomp}
We compare the known travel time of each trip in the test data with the
point and 95\% interval predictions from each method. Unlike the
simulated test data in Section \ref{secsimexp}, the true paths are
not known. For the Bayesian and local methods, we assume that the path
taken is the fastest path in expectation. This measures the ability of
each method to estimate both the fastest path and the travel time distributions.

We again use the cross-validation approach of Section \ref{secbias} to
estimate bias correction factors. We repeat this five times, resampling
random training and validation/test sets, and give arithmetic means of
the performance metrics over the five replications in Table \ref
{taboutofsample}. We again compare the point estimates from the
three methods on the test data using RMSE, RMSE log, and Bias (M.A.),
and compare the interval estimates using Width and Cov. \%. Because
the true travel time distributions are unknown, we cannot use the
Oracle method as in Section \ref{secsimpredict}. However, we still
wish to estimate gold standard performance, so we implement an
Estimated Oracle method, in which we assume that the parametric model
and MLE estimates from the Local MLE method are the truth. We simulate
realized travel times on the fastest path (in expectation, as estimated
by the Local MLE method) for each test trip and compare these to the
point estimates from the Local MLE method. To avoid simulation error,
we use Monte Carlo estimates from 1000 simulated travel times for each trip.

\begin{table}
\caption{Out-of-sample trip travel time estimation performance on
Toronto EMS data}
\label{taboutofsample}
\begin{tabular*}{\textwidth}{@{\extracolsep{\fill}}lccccc@{}}
\hline
\textbf{Estimation method} & \textbf{RMSE (s)} & \textbf{RMSE log} & \textbf{Bias (M.A.)} & \textbf{Cov. \%} &
\textbf{Width (s)} \\
\hline
\multicolumn{6}{c}{L--S data (Mean over five replications)} \\
Est. oracle & \phantom{0}14.9 & 0.168 & 0.018 & -- & -- \\
Bayesian & \phantom{0}37.8 & 0.332 & 0.025 & 85.8 & \phantom{0}75.0 \\
Local MLE & \phantom{0}38.4 & 0.342 & 0.027 & 73.3 & \phantom{0}55.0 \\
Local harm. & \phantom{0}38.5 & 0.343 & 0.028 & 77.5 & \phantom{0}75.2 \\
Budge et al. & \phantom{0}39.8 & 0.342 & 0.028 & 94.5 & 122.3 \\
\multicolumn{6}{c}{Std data (Mean over five replications)} \\
Est. oracle & \phantom{0}35.2 & 0.191 & 0.018 & -- & -- \\
Bayesian & 126.8 & 0.465 & 0.025 & 73.0 & 141.8 \\
Local MLE & 129.0 & 0.480 & 0.025 & 58.4 & 118.6 \\
Local harm. & 129.0 & 0.480 & 0.025 & 64.8 & 142.8 \\
Budge et al. & 127.9 & 0.475 & 0.026 & 94.3 & 370.8 \\
\hline
\end{tabular*}
\end{table}

For the L--S data, the Bayesian method outperforms the method of Budge
et al. and the local methods, suggesting that it is effectively
combining trip information with GPS information. The Bayesian method is
roughly 6\% better in log scale RMSE, after subtracting the error from
the Estimated Oracle method. The method of Budge et al. and the local
methods perform similarly. The bias correction is successful at
eliminating bias (there is 2--3\% bias remaining).

The Bayesian method substantially outperforms the local methods in
interval estimates. The Bayesian intervals have much higher coverage
percentage than the intervals from the local methods. The method of
Budge et al. has higher coverage percentage than the Bayesian method,
however, the intervals are also wider.\vadjust{\goodbreak} The intervals from the MLE
method are narrow and have low coverage percentage. Therefore, the
Local MLE method does not adequately account for travel time
variability, suggesting that the Estimated Oracle method may
underestimate the baseline error. If so, the Bayesian method
outperforms the other methods by an even larger amount, relative to the
baseline error.

For the Std data, the Bayesian method outperforms the local methods by
roughly 5\% in RMSE on the log scale, and outperforms the method of
Budge et al. by 3.5\%, again relative to the Estimated Oracle error.
Point estimates from the method of Budge et al. slightly outperform
the local methods. Interval estimation is less successful for the
Bayesian and local methods than for the L--S data, probably because the
Std travel times have more unaccounted sources of variability than the
L--S travel times, such as traffic and time of day.

This region and data set are generally favorable to the method of Budge
et al. The travel speeds are similar across most roads in this region,
which mitigates the main weakness of the Budge et al. method, namely,
its inability to distinguish between fast and slow roads. Also, several
particular paths are very common in the Leaside region, and the Budge
et al. method fits the travel time distribution of these particular
paths very closely, leading to relatively high predictive accuracy. On
the full city the routes would be much more heterogeneous, with many
different routes of roughly the same travel distance, so that a method
that can model the heterogeneity is expected to have a greater advantage.

\subsection{Response within time threshold}
\label{secthresh}
\begin{figure}

\includegraphics{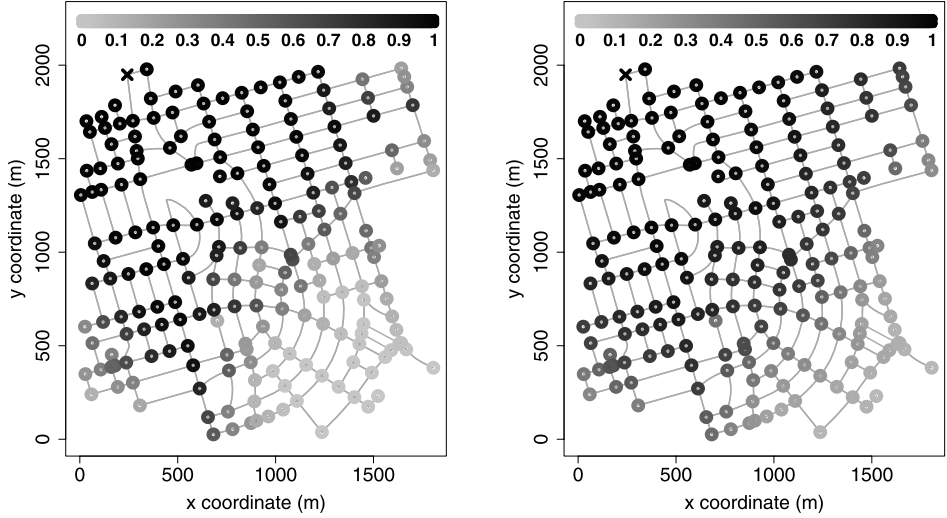}

\caption{Estimates of probability of reaching each node in 150 seconds,
Bayesian method (left), Budge et al. method (right), from the location
marked X.}
\label{fignodereach}
\end{figure}

Next we estimate the probability an ambulance completes its trip within
a certain time threshold [\citet{budge10}]. These probabilities are
critical for EMS providers (see Section \ref{secintro}). In Figure \ref
{fignodereach}, we assume that an ambulance begins at the node marked
with a black~X and estimate the probability it reaches each other node
in 150 seconds, following the fastest path in expectation. For the
Bayesian method, these probabilities are calculated by simulating
travel times from the posterior distribution of each arc in the route,
and using Monte Carlo estimation (see Section \ref{secgibbs}). The
left-hand figure shows probabilities from the Bayesian method, and the
right-hand figure shows probabilities from the method of Budge et al.

\begin{figure}[b]

\includegraphics{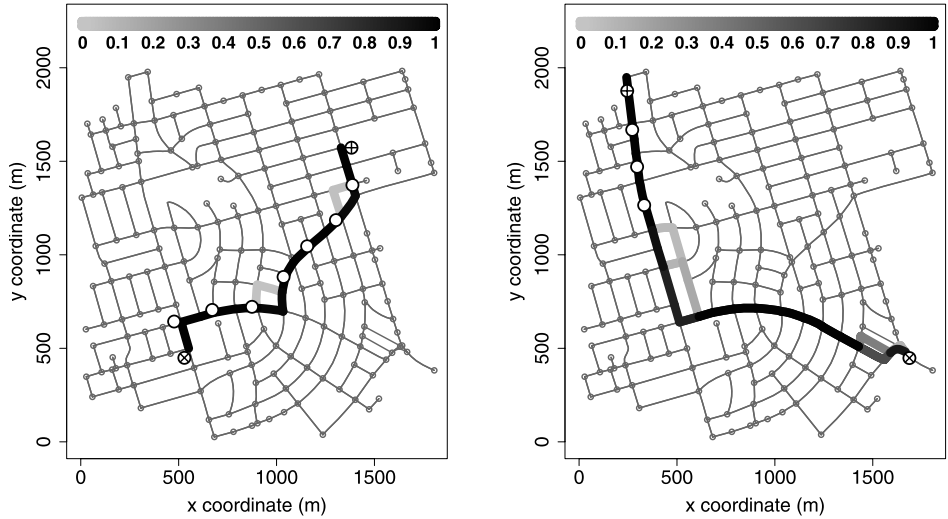}

\caption{Map-matching estimates for two Toronto L--S trips, shaded by
the probability each arc is traversed.}
\label{figrealpost}
\end{figure}

The probabilities for both methods appear reasonable; they are high for
nodes close to the start node and decrease for nodes further away. The
probabilities from the Bayesian method appear more realistic than those
from Budge et al., since nodes on main roads tend to have higher
probabilities from the Bayesian method (e.g., traveling south
from the start node), whereas nodes on minor roads far from the start
node have lower probabilities from the Bayesian method (see the
bottom-right in each plot). This is because the method of Budge et al. does not take into account the different speeds of different roads.

\subsection{Map-matched path accuracy}
\label{secrealmapmatch}
Finally, we assess map-matching estimates from the Bayesian method, for
the Toronto L--S data. Figure \ref{figrealpost} shows two example
ambulance paths from the L--S data set. The GPS locations are shown in
white; the first reading is marked with a cross and the last with an X.
As in Section \ref{secsimmapmatch}, each arc is shaded by its
marginal posterior probability, if it is greater than 1\%. In the
left-hand path, there are two occasions where the path is not precisely
defined by the GPS readings. On both occasions, roughly 90\% of the
posterior probability is given to a route following the main road,
which is estimated to be faster. The final two GPS readings appear to
have location error. However, the fastest path is still given roughly
100\% posterior probability, instead of a detour that would be slightly
closer to the second-to-last GPS reading. In the right-hand path, for
an unknown reason, there is a large gap between GPS points. Most of the
posterior probability is given to the fastest route along the main
roads. This illustrates the robustness of the Bayesian method to sparse
GPS data.

\section{Conclusions}
\label{secconc}
We proposed a Bayesian method to estimate the travel time distribution
on any route in a road network using sparse and error-prone GPS data.
We simultaneously estimated the vehicle paths and the parameters of the
travel time distributions. We also introduced two local methods based
on mapping each GPS reading to the nearest road segment. The first
method used the harmonic mean of the GPS speeds; the second performed
maximum-likelihood estimation for a lognormal distribution of travel
speeds on each segment.

We compared these three methods to an existing method from \citet
{budge10}. In simulations, the Bayesian method greatly outperformed the
local methods and the method of Budge et al. in estimating
out-of-sample trip travel times, for both point and interval estimates.
The estimates from the Bayesian method remained excellent even when the
GPS data had high error. On the Toronto EMS data, the Bayesian method
again outperformed the competing methods in out-of-sample prediction
and provided more realistic estimates of the probability of completing
a trip within a time threshold than the method of Budge et al.

We plan to extend the Bayesian method to include time-varying travel
times. For instance, speeds typically decrease during rush hour.
Applying the methods of this paper separately to rush hour and nonrush
hour improves performance on standard travel Toronto data, although it
has little effect on performance for lights-and-sirens data. A more
sophisticated approach that smooths across time of day may have better success.

We are currently investigating a number of other extensions. First, we
are developing methods to approximate or modify the Bayesian method to
obtain efficient computation on very large networks. Second, we are
experimenting with information sharing across roads to improve
estimates on infrequently used roads. Third, we are incorporating
dependence between arc travel times within each trip, arising from
traffic congestion effects or a driver's speed preference, for example.
This change is expected to improve coverage of interval estimates.
Finally, we are investigating the use of turn penalties. For example, a
left turn can require more time than a right turn.

\section*{Acknowledgments}
We thank the referees and Associate Editor for their careful reading
and comments. We also thank The Optima Corporation and Dave Lyons of
Toronto EMS for their collaboration.

\begin{supplement}[id=suppA]
\stitle{Appendix A, B, C}
\slink[doi]{10.1214/13-AOAS626SUPP} 
\sdatatype{.pdf}
\sfilename{aoas626\_supp.pdf}
\sdescription{Appendix A: Constants and hyperparameters.
Appendix B: Reversibility of the path update.
Appendix C: Harmonic mean speed and
GPS sampling.}
\end{supplement}

%

%

\printaddresses

\end{document}